\documentclass[fleqn,twoside]{article}
\usepackage{gc,cite,graphicx,epsf,amssymb}

\def\thd{\fract 13}

\def\kappa{\varkappa}

\def\mn{_{\mu\nu}}

\def\mN{_\mu^\nu}

\def\R{{\mathbb R}}
\def\Z{{\mathbb Z}}
\def\dens{\mbox{${\rm g\,\cdot\,cm}^{-3}$}}

\def\GR{general relativity}

\def\sph{spherically symmetric}
\def\ssph{static, spherically symmetric}

\def\bh{black hole}
\def\bhs{black holes}
\def\wh{wormhole}
\def\whs{wormholes}
\def\bw{brane world}
\def\bwd{brane-world}
\def\asflat{asymptotically flat}

\def\KS{Kantowski-Sachs}

\bls{1.04}
\begin{document}
\twocolumn[
\prepno{yymm.nnnn}{}

\Title {Possible black universes in a brane world}

\Aunames{K.A. Bronnikov\auth{a,b,1} and E.V. Donskoy\auth{b,2}}

\Addresses{
\addr a {Center of Gravitation and Fundamental Metrology,
         VNIIMS, Ozyornaya St. 46, Moscow 117361, Russia}
\addr b {Institute of Gravitation and Cosmology,
         PFUR, Miklukho-Maklaya St. 6, Moscow 117198, Russia}
          }

\bigskip
%\Rec{}

\Abstract
{A black universe is a nonsingular black hole where, beyond the horizon,
 there is an expanding, asymptotically isotropic universe. Such spherically
 symmetric configurations have been recently found as solutions to the
 Einstein equations with phantom scalar fields (with negative kinetic energy)
 as sources of gravity. They have a Schwarzschild-like causal structure but a
 de Sitter infinity instead of a singularity. It is attempted to obtain
 similar configurations without phantoms, in the framework of an RS2 type
 brane world scenario, considering the modified Einstein equations that
 describe gravity on the brane. By building an explicit example, it is shown
 that black-universe solutions can be obtained there in the presence of a
 scalar field with positive kinetic energy and a nonzero potential.}

] %%%%%%%%%%%%%%%%%%%%%%%%%%%%%%%%%%%%%%%
\email 1 {kb20@yandex.ru}
\email 2 {cosmit3000@yandex.ru}

\section{Introduction}

  The problem of singularities is one of the long-standing problems in
  the classical theories of gravity. Singularities are places where general
  relativity or another classical theory of gravity does not work. Therefore,
  a full understanding of the physics of phenomena under study (origin and
  fate of our Universe, gravitational collapse etc.) requires avoidance of
  singularities or/and modification of the corresponding classical theory or
  addressing quantum effects. There have been numerous attempts on this
  trend, some of them suggesting that singularities inside the event horizons
  of \bhs\ should be replaced with a kind of regular core (\cite{dym92}, see
  \cite{dym06} for a recent review), others describing bouncing or
  ``emergent'' universes (see, e.g., \cite{bounce, emergent} for reviews).

  In our view, of particular interest are models which combine avoidance
  of singularities in both \bhs\ and cosmology, those which have been termed
  {\it black universes\/} \cite{pha1, pha4}. These are regular \sph\ \bhs,
  with the same causal structure as the Schwarzschild \bh, but where a
  possible explorer, after crossing the event horizon, gets into an expanding
  universe instead of a singularity. Thus such hypothetic configurations
  combine the properties of a wormhole (absence of a center, a regular
  minimum of the area function) and a \bh\ (a Killing horizon separating R
  and T regions). Moreover, the \KS\ cosmology in the T region is
  asymptotically isotropic and approaches a de Sitter regime of expansion,
  which makes such models potentially viable as models of our accelerating
  Universe.

  Such objects are naturally obtained \cite{pha1, pha4} if one considers local
  concentrations of dark energy (DE) represented by different forms of phantom
  matter: phantom scalars in scalar-tensor theories of gravity and the
  so-called k-essence whose most general Lagrangians have the form
  $F(\phi, X)$ where $X = \nabla_\alpha\phi \nabla^\alpha\phi$. In all such
  cases, there is at least one classical field with a negative kinetic term.

  The phantom behavior of DE (such that its pressure to density ratio
  $w = p/\eps < -1$) is favored by modern cosmological observations. More
  precisely, the ``Gold'' supernova sample data \cite{R04} slightly favor a
  phantom behavior of DE at small redshifts $z < 0.3$ along with crossing the
  phantom divide $w=-1$ at larger $z$ \cite{phantom04} (see \cite{SS06} for
  further references). Other cosmological data suggest a cosmological
  constant as the best fit but still do not exclude recent phantom DE
  behavior, see \cite{phantom06}. The latest estimates of $w$ also peak
  somewhat near $-1$ and admit $w<-1$ \cite{w09}.

  Meanwhile, there exist models that admit a phantom DE behavior without
  explicitly introducing phantom fields. Among them the simplest is the
  generic scalar-tensor gravity with non-zero scalar field potentials
  (see, e.g., \cite{GPRS06}) which has sufficient freedom to describe all
  observational data.

  There are theoretical reasons for considering phantom fields: they
  naturally appear in some models of string theory \cite{sen}, supergravities
  \cite{sugr} and theories in more than 11 dimensions like F-theory
  \cite{khvie}.

  Nevertheless, one should bear in mind that a classical field with a negative
  kinetic term can have an arbitrarily large negative energy of high-frequency
  oscillations, which is quite undesirable from the viewpoint of quantum
  field theory: it can lead to runaway production of particles and
  antiparticles accompanied by production of equal negative energy of the
  phantom field itself (see, e.g., \cite{CGM04}). Nothing of this kind is
  observed, which casts serious doubt on possible existence of phantom fields.
  Moreover, as was recently argued in \cite{GPRS06}, cosmological models with
  a phantom scalar field cannot explain the observed large-scale homogeneity
  and isotropy of the Universe.

  Thus there exist arguments both {\it pro et contra\/} phantom fields, and
  the latter seem somewhat stronger. In any case, it is reasonable to try to
  avoid such fields in modelling real or hypothetic phenomena.

  Accordingly, we here try to show that black-universe models can be obtained
  without invoking phantom fields. This appears to be possible in the
  framework of the \bw\ scenario, using the modified Einstein equations
  \cite{SMS99} describing gravity on the brane.

  The \bw\ concept describes our world as a 4D surface (brane) supporting
  all or almost all matter fields and embedded in a higher-dimensional
  space-time (called the bulk). This concept traces back to the early 80s
  and leads to a variety of consequences in cosmology, gravitational and
  particle physics, see the reviews \cite{bra-rev}.
  In particular, brane worlds turn out to be a natural framework for \whs\
  \cite{bwh1, camera} (see also references therein) since there the
  modified Einstein equations \cite{SMS99} [see \eqs (\ref{EE4})] contain a
  source term $E\mN$ of geometric origin which need not observe the usual
  energy conditions. And, as we shall see in the present paper, it is this
  source term that can replace phantom fields in building black-universe
  models.

  The modified Einstein equations (EE4) used here correspond to the so-called
  RS2 scenario: a single brane in a $\Z_2$-symmetric 5-dimensional bulk, with
  all fields except gravity confined on the brane. It generalizes the second
  Randall-Sundrum model comprising a single Minkowski brane in an anti-de
  Sitter (AdS) bulk \cite{RS2}. However, in other brane-world scenarios, the
  effective 4D Einstein equations also contain terms similar to $E\mN$, e.g.,
  on codimension-1 branes without $Z_2$ symmetry \cite{YaSa07} and in
  Dvali-Gabadadze-Porrati brane worlds \cite{DGP} with different kinds of
  induced gravity terms \cite{Saav08}. Thus we can anticipate that
  black-universe solutions similar to ours exist in such brane worlds as
  well, though probably under some other conditions.

  The paper is organized as follows. In Section 2 we briefly discuss the
  Einstein-scalar field equations for \ssph\ systems in \GR\ and present the
  simplest black-universe solution with a phantom scalar. In Section 3
  we analyze the similar set of equations in a \bw\ and show that for our
  purpose we can neglect some terms, namely, those quadratic with respect to
  the stress-energy tensor of matter. Section 4 is devoted to attempts to
  obtain black-universe solutions by properly employing the freedom that
  exists in this system. Section 5 contains some concluding remarks.

\section{Black-universe solutions in \GR}  \label{s2}

  Consider the general \ssph\ metric
\beq                                                            \label{ds}
    ds^2 = A(u) dt^2 - \frac{du^2}{A(u)} - r^2(u)d\Omega^2,
\eeq
  where $d\Omega^2 = d\theta^2 + \sin^2\theta d\varphi^2$ is the metric on
  a unit sphere.\footnote
    {Our sign conventions are as follows: the metric signature
         $(+{}-{}-{}-)$; the curvature tensor $R^{\sigma}{}_{\mu\rho\nu} =
         \d_\nu\Gamma^{\sigma}_{\mu\rho}-\ldots$, so that, e.g., the Ricci
         scalar $R > 0$ for de Sitter space-time, and the stress-energy
         tensor (SET) such that $T^t_t$ is the energy density.}
  The metric (\ref{ds}) is written in terms of the ``quasiglobal'' coordinate
  $u$, which is particularly convenient for dealing with Killing horizons
  where it behaves in the same way as the manifestly well-behaved
  Kruskal-like null coordinates. For this reason, in terms of $u$, one
  may consider regions on both sides of such a horizon remaining in a
  formally static framework.

  The two functions, $A(u)$ (often called the redshift function) and
  $r(u)$ (the area function, equal to the radius of a coordinate sphere
  at given $u$) entirely determine the geometry under consideration.
  Horizons correspond to regular zeros of $A(u)$.

  Our interest is to find black-universe solutions which, by definition,
  must have the following properties:
\begin{enumerate} \itemsep 1pt
\item
     Regularity in the whole range $u \in \R$;
\item
     Asymptotic flatness as $u \to -\infty$ (without loss of generality),
     i.e., $r(u) \approx -u$, $A(u)\to 1$;
\item
     A de Sitter asymptotic as $u \to +\infty$, i.e., a T region
     ($A < 0$) where $r(u)\sim u$, $-A(u)\sim u^2$;
\item
     A single simple horizon (i.e., a simple zero of $A(u)$) at finite $u$.
\end{enumerate}

  As shown in \cite{pha4}, such solutions can exist in \GR\ with various
  phantom sources. Let us here present the simplest example with a minimally
  coupled scalar field having the Lagrangian
\beq                                                          \label{L_s}
     L_{\phi} = \Half \eps\d^\alpha\phi \d^\alpha\phi - V(\phi),
\eeq
  where $\eps=+1$ corresponds to normal scalar fields with positive kinetic
  energy, $\eps=-1$ to phantom fields, and $V(\phi)$ is a potential. The set
  of Einstein-scalar equations for the metric (\ref{ds}) and
  $\phi=\phi(u)$ may be written in the form
\bear
    \eps (Ar^2 \phi')' \eql - r^2 dV/d\phi,               \label{phi''}
\\
              (A'r^2)' \eql - 2r^2 V;                         \label{00}
\\
              2 r''/r \eql \eps\phi'{}^2 ;                    \label{01}
\\
         A (r^2)'' - r^2 A'' \eql 2 ,                         \label{02}
\ear
  where the prime denotes $d/du$. The scalar field equation (\ref{phi''})
  is a consequence of \eqs (\ref{00})--(\ref{02}), which, given a potential
  $V(\phi)$, form a determined set of equations for the unknowns
  $r(u),\ A(u),\ \phi(u)$.

  As is evident from (\ref{01}), black-universe solutions cannot be obtained
  with $\eps=+1$ because in this case $r''\leq 0$ which is incompatible with
  requirements 1-3 (instead of $u\in \R$, there will be inevitably $r=0$
  at some finite $u$, which is either a singularity or, at best, a regular
  center).

  A particular solution to these equations with $\eps=-1$ is given by
  \cite{pha1}
\bear                                                         \label{r1}
    r \eql (u^2 + b^2)^{1/2}, \cm b = \const > 0;
\yy                                                           \label{B1}
       B(u) \eqv \frac{A(u)}{r^2(u)}
      = \frac{c}{b^2} + \frac{1}{b^2+u^2}
\nnn \cm
      + \frac{u_0}{b^3}
    \left(\frac{bu}{b^2 + u^2}
            + \arctan \frac{u}{b}\right),
\yy                                                           \label{phi1}
      \phi \eql \pm\sqrt{2} \arctan (u/b) + \phi_0,
\yy                                                           \label{V1}
     V \eql - \frac{c}{b^2} \frac{r^2 + 2u^2}{r^2}
\nnn \ \ \
        - \frac{u_0}{b^3} \left( \frac{3bu}{r^2}
            + \frac{r^2 + 2u^2}{r^2}\arctan \frac{u}{b}\right)
\ear
   with $c,\ u_0,\ \phi_0 =\const$. In particular,
\beq                                                       \label{BV_as}
      B(\pm \infty) = -\frac 13 V(\pm \infty)
            = \frac{2bc \pm \pi u_0}{2b^3}.
\eeq
  Choosing in (\ref{phi1}), without loss of generality, the plus sign and
  $\phi_0=0$, we obtain for $V(\phi)$
\bearr
     V(\phi) = -\frac{c}{b^2} (3 - 2\cos^2 \psi)             \label{Vf}
\nnn \cm
     - \frac{u_0}{b^3} \left[3\sin\psi \cos\psi
                    + \psi (3 - 2\cos^2 \psi) \right],
\nnn
  \psi := \frac{\phi}{\sqrt{2}}.
\ear

  The solution is \asflat\ at large negative $u$ ($B\to 0$, $A\to 1$)
  under the condition $2bc = \pi u_0$, and the Schwarzschild mass, defined
  in the usual way, is then $m = -u_0/3$. Then, assuming $m >0$,
  we find that $B(+\infty) = -3\pi m/b^3 = \const <0$, which corresponds to
  a de Sitter asymptotic with a cosmological constant $\Lambda >0$. The
  horizon position is found by solving the transcendental equation
  $B(u)=0$.

  This is an example of a black-universe solution. Other cases of the
  solution (\ref{r1})--(\ref{V1}) include wormholes and asymptotically AdS
  configurations, see more details in \cite{pha1, pha4}.

\section{Gravity on the brane}    \label{s3}

\subsection{Modified Einstein equations}

  The gravitational field on the brane is described by
  the modified Einstein equations \cite{SMS99}
\bear
    G\mN = - \Lambda_4\delta\mN -\kappa_4^2 T\mN
            - \kappa_5^4 \Pi\mN - E\mN,                     \label{EE4}
\ear
  where $G\mN = R\mN - \half \delta\mN R$ is the 4D Einstein tensor,
  $\Lambda_4$ is the 4D cosmological constant expressed in terms of
  $\Lambda_5$ and the brane tension $\lambda$:
\beq
    \Lambda_4 = \Half
    \biggl(\Lambda_5 + \frac{1}{6} \kappa_5^4\lambda^2\biggr);  \label{La4}
\eeq
  $\kappa_4^2 = 8\pi G_N = \kappa_5^4 \lambda/(6\pi) = m_4^{-2}$ is the 4D
  gravitational constant; $G_N$ is Newton's constant of gravity, and $m_4$
  is the 4D Planck mass;

\medskip\noi
  $\kappa_5 = m_5^{-3/2}$, $m_5$ being the 5D Planck energy scale;

\medskip\noi
  $T\mN$ is the SET of matter trapped on the brane;

\medskip\noi
  $\Pi\mN$ is a tensor quadratic in $T\mN$, obtained from matching the
  5D metric across the brane:
\beq  \nq                                                     \label{Pi_}
    \Pi\mN = \fract{1}{4} T_\mu^\alpha T_\alpha^\nu - \half T T\mN
           - \fract{1}{8} \delta\mN
    \left( T_{\alpha\beta} T^{\alpha\beta} -\thd T^2\right),
\eeq
  where $T = T^\alpha_\alpha$; lastly, $E\mN$ is the ``electric'' part of
  the 5D Weyl tensor projected onto the brane: in proper 5D coordinates,
\beq                                                             \label{E_}
    E\mn = \delta_\mu^A \delta_\nu^C {}^{(5)} C_{ABCD} n^B n^D
\eeq
  where $A, B, \ldots$ are 5D indices and $n^A$ is the unit normal to the
  brane. By construction, $E\mN$ is traceless, $E_\mu^\mu = 0$.

  Other characteristics of $E\mN$ are unknown without specifying the 5D
  metric, hence the set of equations (\ref{EE4}) is not closed. In isotropic
  cosmology this leads to an additional arbitrary constant in the field
  equations, connected with the density of ``dark radiation'' \cite{bra-rev}.
  For \ssph\ systems to be discussed here, this freedom is expressed in the
  existence of one arbitrary function of the radial coordinate.

\subsection{Reasons for neglecting $\Pi\mN$}

  Let us show that under quite reasonable conditions we can neglect the
  tensor $\Pi\mN$ in (\ref{EE4}).

  We put $\Lambda_4 =0$, so that
\beq
      |\Lambda_5| = \frac{1}{6} \kappa_5^4 \lambda^2
              = 6\pi^2 (\kappa_4/\kappa_5)^4,              \label{La5}
\eeq
  and use the observational restriction on the bulk length scale $\ell$
  which follows from the recent short-range Newtonian gravity tests
  \cite{Newt}, showing that Newton's inverse-square law hold at length
  scales greater than about 0.1 mm. This means that if we live on an
  RS2-like brane, the bulk length scale can be estimated as
\beq
    \ell = (6/|\Lambda_5|)^{1/2} \lesssim 10^{-2}\ {\rm cm}.  \label{ell}
\eeq
  Note that the 4D Planck scale in our notation is
\bearr
      m_4 = \kappa_4^{-1} = (8\pi G_{\rm N} \approx 2.4\ten{18}\,{\rm GeV},
\nnn
      l_4 = 1/m_4 = \kappa_4 \approx 8\ten{-33}\,{\rm cm}.
\earn
  Combining (\ref{La5}) and (\ref{ell}), we obtain
\beq                                                          \label{m5}
       m_5/m_4 = (\pi \ell/l_4)^{-1/3} \gtrsim 10^{-10},
\eeq
  so that the 5D Planck energy scale in this scenario is at least about
  $10^8$ GeV.

  Now, we can assert that the term with $\Pi\mN$ is negligible in (\ref{EE4})
  as compared with the $T\mN$ term as long as
\[
      \kappa_5^4 W^2 \ll \kappa_4^2 W,
\]
  (where $W$ characterizes the magnitude of $T\mN$, say, the absolute
  value of the largest component of $T\mN$), or
\beq
       W \ll m_5^6/m_4^2 = m_4^4 (m_5/m_4)^6,           \label{W<}
\eeq
  where $m_4^4 \approx 3.5\ten{73} {\rm GeV}^4 \approx 8.4\ten{90}\ \dens$ is
  the Planck density while the second factor is, according to the
  experimental bound (\ref{m5}), about $10^{-60}$ or larger. As a result, for
  the ``density'' $W$ we have
\[
        W  \ll 10^{30}\ \dens.
\]
  Recalling that the density of nuclear matter is about $10^{13}$ \dens, it
  is clear that this bound certainly holds for any thinkable matter.

\subsection{Brane gravity with a scalar field}

  Consider \eqs (\ref{EE4}) for \ssph\ configurations of a normal scalar
  field with an arbitrary potential, neglecting the term $\Pi\mN$. So the
  scalar field Lagrangian has the form (\ref{L_s}) with $\eps=1$. The
  general \ssph\ metric is again taken in the form (\ref{ds}), with the
  quasiglobal coordinate $u$.

  The scalar field EMT is conservative, so the same is required for
  $E\mN$. If we take it, for convenience, in the form
\bearr      \nq\,
    E\mN = \diag (-P{-}Af,\ -P,\ P{+}Af/2,\ P{+}Af/2),     \label{E_mn}
\nnn
\ear
  where $P$ and $f$ are some functions of the radial coordinate $u$
  (so that, as required, its trace is zero), then the conservation law
  $\nabla_\alpha E^\alpha_1$ is written as
\beq
    (Pr^4)' = \half r^6 f (A/r^2)'.                     \label{E-cons}
\eeq

    \eqs (\ref{EE4}) may be written in the form
\bear
            \frac{1}{2r^2}(A'r^2)' \eql - V - P - Af;           \label{00}
\\
              2 r''/r \eql - {\phi'}^2 + f ;                    \label{01}
\\
         A (r^2)'' - r^2 A'' - 2 \eql 2P + \frac{3}{2}Af.       \label{02}
\ear
  The scalar field equation $(Ar^2 \phi')' = r^2 dV/d\phi$ follows from
  (\ref{00})--(\ref{02}) combined with (\ref{E-cons}).

  Thus, if $V(\phi)$ is specified, we have four independent equations
  (\ref{E-cons})--(\ref{02}) for five unknown functions of $u$:
  $\phi,\ A,\ r,\ f$ and $P$. The system becomes still more underdetermined
  if $V(\phi)$ is not specified: in this case we can choose as many as two
  functions by hand to obtain a solution.

\section {Attempts to obtain black-universe solutions}  \label{s4}

\subsection{Models with $r'' > 0$}

  To have a regular positive minimum of $r(u)$, we must have $r''>0$ at least
  in some range of $u$. By (\ref{01}), this can be achieved only with $f >
  0$. We first try to obtain such a solution with $r'' >0$ {\it everywhere\/},
  i.e., $f > \phi'{}^2$.

  To integrate (\ref{E-cons}), let us suppose
\beq                                                           \label{f1}
     f=2C/r^6, \cm C=\const.
\eeq
     Then we have
\bearr                                                           \label{P1}
    P r^4 = CB + C_1, \cm B(u) \equiv A/r^2,
\nnn
    \cm C_1 = \const.
\ear
  Putting for simplicity $C_1=0$ and substituting (\ref{P1}) into
  (\ref{02}), we obtain an equation connecting $B(u)$ and $r(u)$:
\beq
     (r^4 B')' + 2 + 10 CB/r^2 =0.                            \label{B''}
\eeq
  If both $r$ and $B$ are known, the quantities $\phi$ and $V$ are found
  from (\ref{00}) and (\ref{01}). Thus it remains to choose $r$ and $B$
  satisfying (\ref{B''}) and providing the above properties 1--4.

  This task turns out to be hard if at all solvable. Thus, if we choose
  $r^2 = u^2 + b^2$ ($b=\const >0$) with good asymptotics at large $u$, we
  obtain $r''/r = b^2/r^4$, whence (\ref{01}) combined with (\ref{f1}) leads
  to $\phi'{}^2 < 0$ at large $u$, i.e., a contradiction.

  Another choice, $r^4 = u^4 + b^4$, is compatible with (\ref{01}) if
  $C > 3b^4$. However, a numerical solution of \eq (\ref{B''}) leads to
  functions $B(u)$ with at least 3 zeros, in other words, at least 3 horizons
  in space-time (see an example in Fig. 1).

\begin{figure}
    \begin{center}
\fbox{\includegraphics[width=8.2cm,height=5cm]{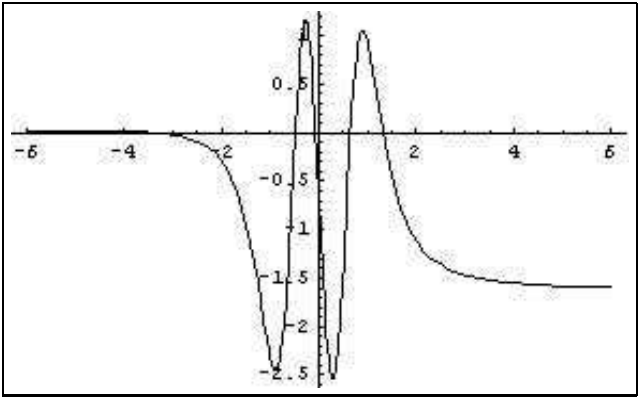}}
    \end{center}
\caption{Example of an oscillating solution for $B(x)$}
\hrulefill\                                                 %% fig1
\end{figure}

  The nature of \eq (\ref{B''}) apparently implies an oscillatory behavior
  of $B(u)$ if the coefficient $10 C/r^2$ (originating from $E\mN$) is large
  enough in a sufficiently wide range of $u$. A possible way of avoiding
  such a behavior is to try a ``more concentrated'' distribution of $f$ and
  $P$.

\subsection{Models with $f \sim \delta(u)$}

  Let us try to find a black-universe solution assuming that $f(u)$ is
  concentrated on a single sphere, e.g., $u=0$.

  We now make the equations dimensionless by putting
\bearr                                                          \label{dim-}
      x = u/b, \ \ \ Pr^4 = Q(x)b^2, \ \ \ fr^4 = 2F(x)b^2,
\nnn
      r = b{\bar r}(x), \ \ \  B = {\bar B}(x)/b^2,
\ear
  where $b = \const >0$ specifies a length scale of the configuration, and to
  justify our rejection of $\Pi\mN$ we assume $b \gg \ell$ (see above).

  In what follows we omit the bars over $r$ and $B$.
  \eqs (\ref{E-cons}), (\ref{01}) and (\ref{02}) take the form
\bearr
        Q' = r^2 F B',                      \label{cons}
\yyy
    \frac{r''}{r} = -\psi'{}^2 + \frac{F}{r^4},     \label{01'}
\yyy
    (r^4 B')' + 2 + \frac{4Q}{r^2} + 6 BF =0,       \label{02'}
\ear
  where the prime denotes $d/dx$ and $\psi = \phi/\sqrt{2}$.

  To begin with, we choose $r(x)$ so that $r'' < 0$ at all $x\ne 0$:
\beq
    r^2(x) = (|x| + c)^2 -1, \cm   c = \const > 1.   \label{r(x)}
\eeq
  This conforms to both flat and de Sitter asymptotics. Then, in \eq
  (\ref{01'}), the quantity $r''/r$ has a delta-like singularity at $x=0$.
  To compensate it and make $\psi'$ continuous at $x=0$, we put
\beq
    F(x) = 2c r_0^2 \delta(x), \cm r_0 = \sqrt{c^2 -1}.  \label{F(x)}
\eeq
  We have then
\beq
    \psi' = \pm \frac{1}{r^2} = \frac{1}{(|x| +c)^2 - 1}, \label{psi'}
\eeq
  whence, without loss of generality,
\bearr
    \psi = \vars{ h(c-x), & x<0,\\
              2h(c) - h(c+x), & x >0,                \label{psi}
              }
\nnn
    h(x) := \Half \ln \left| \frac{x+1}{x-1} \right|.
\ear
     \eq (\ref{cons}) gives for $Q(x)$:
\beq
    Q(x) = Q_0 + 2c r_0^4 B_{x0} \theta(x),
\eeq
  where $Q_0 = \const$, $B_{x0} = B'(0)$ and
\[
  \theta(x) = \vars {1, & x>0, \\ 0, & x<0.}
\]
  is the Heaviside function.

\begin{figure}
    \begin{center}
\fbox{\includegraphics[ width=8.2cm,height=8cm]{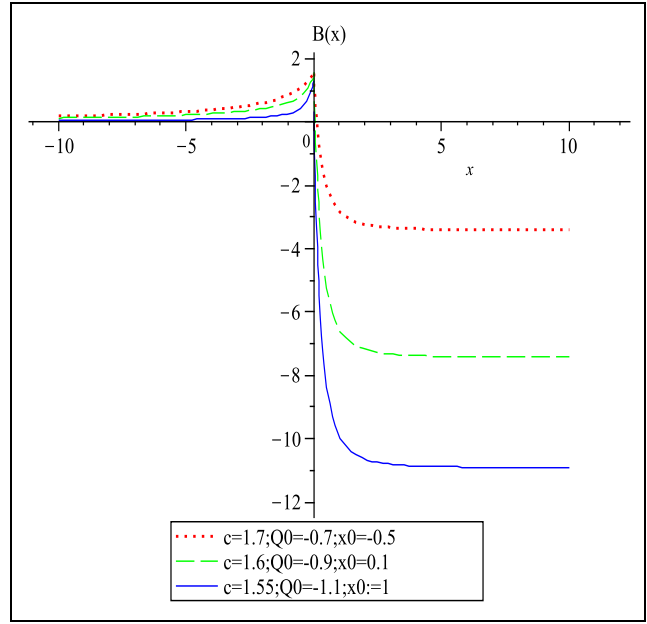}}
    \end{center}
\caption{The function $B(x)$ obtained from a delta-like $f(u)$}
\hrulefill\
\end{figure}   % fig2

  It remains to solve \eq (\ref{02'}) and to find $V$. \eq (\ref{02'}) now
  has the form
\bearr
      (r^4 B')' = -2 - \frac{4Q_0}{r^2}                       \label{B_xx}
            - \frac{8cr_0^4 B_{x0}}{r^2} \theta(x)
\nnn \inch
            - 12 B_0 cr_0^2 \delta(x),
\ear
  where $B_0 = B(0)$. We solve (\ref{B_xx}), choosing the initial
  conditions at large negative $x$ in the Schwarzschild form
\beq
       A(x) = r^2(x) B(x) = 1 - \frac{2m}{|x|},
                \ \ \    m = \const.     \label{ini-}
\eeq
  Integrating (\ref{B_xx}) once, we obtain
\bearr
       r^4 B' = 2(x_0-x) -4Q_0\,h(c-x)                        \label{B_x}
\nnn \nq
            -8 \theta(x) cr_0^4 B_{x0} [h(c) - h(x{+}c)]
                - 12 B_0 c r_0^2 \theta(x),
\ear
  where $x_0 = -m/3$, see (\ref{ini-}). The constant $B_{x0}$ is found as
\beq                                                          \label{B_x0}
    r_0^4 B_{x0} = 2x_0 -4Q_0 h(c) -6 B_0 c r_0^2,
\eeq
  where we have used the principal value of $\theta(0)$ equal to 1/2.
  Note that the term with $B_{x0}$ in (\ref{B_x}) does not contribute
  to the expression (\ref{B_x0}) because it vanishes at $x=0$.

  To find $B(x)$, we first integrate (\ref{B_x}) from $-\infty$ to $x\leq 0$,
  obtaining
\bearr                                                        \label{B-}
       B(x) \Big|_{x \leq 0} =
         \frac{1}{r^2} + (x_0-c)\biggl[\frac{c-x}{r^2(x)} - h(c-x)\biggr]
\nnn \ \
       + Q_0 \biggl[\frac{1-2(c-x)h(c-x)}{r^2(x)} + h^2(c-x)\biggr].
\ear
  This also gives the constant $B_0 = B(0)$,
\bearr                                                          \label{B_0}
       B_0 = \frac{1}{r_0^2} + (x_0-c)\biggl[\frac{c}{r_0^2} - h(c)\biggr]
\nnn \cm
            + Q_0 \biggl[\frac{1-2c h(c)}{r_0^2} + h^2(c)\biggr]
\ear
  which may be used to obtain $B(x)$ at positive $x$ from (\ref{B_x}) by
  integration from 0 to $x > 0$:
\bearr                                                       \label{B+}
       B(x) \Big|_{x \geq 0} = B_0
        + \frac{1}{r_0^2} - \frac{1}{r^2}
\nnn \ \ \
    + [(x_0+c) - 4h(c)(Q_0 + cr_0^4 B_{x0}) - 6 B_0 cr_0^2]
\nnn \inch
       \times \biggl[-\frac{c+x}{r^2} + h(c+x)\biggr]^x_0
\nnn
    + (Q_0 + 2 cr_0^4 B_{x0}) \biggl[\frac{1 - 2(c+x)h(c+x)}{r^2(x)}
\nnn \inch
            + h^2(c{+}x)\biggr]^x_0,
\ear
  where $[f(x)]^b_a := f(b) - f(a)$. This solution (see Fig. 2) really
  describes a black universe, but the delta-like distribution of the
  effective exotic matter, characterized by $f(u)$, causes an undesirable
  discontinuity of $B'$ at $x=0$.

  The expression for the potential $V(x)$ is rather cumbersome and will not
  be presented here.

\subsection{A model with smoothed $f(u)$}

\begin{figure}
     \begin{center}
\fbox{
       \includegraphics[width=3.8cm]{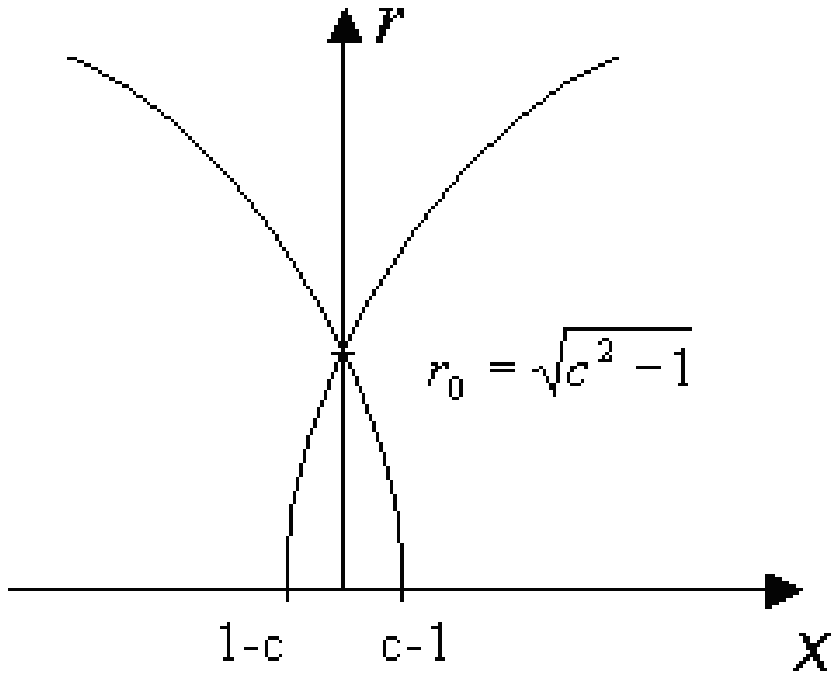}\hfill
        \includegraphics[width=4.3cm]{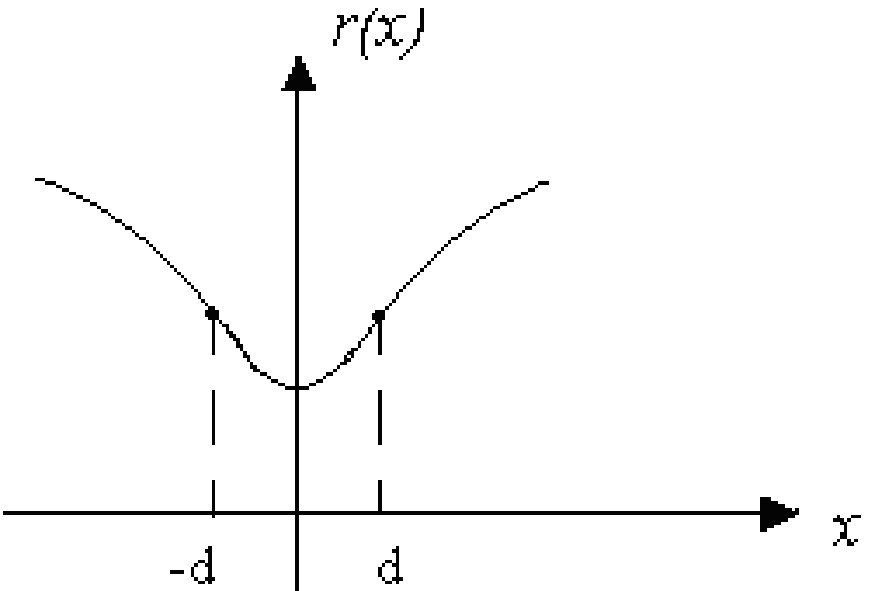}
     }

    \parbox{8cm}{\cm\ \ a \inch\cm  b \cm}
      \end{center}
\caption{The area function $r(x)$ chosen with a delta-like (a)
    and smooth (b) function $f(r)$}
\hrulefill\
\end{figure}    % fig3

  Evidently the qualitative behavior of the model will not change if
  we replace the delta-like distribution of $f(x)$ with a smooth one but
  sufficiently peaked near $x=0$. We will present an example of such a
  solution. Namely, let us preserve the notations (\ref{dim-}), so that the
  field equations have the form (\ref{cons})--(\ref{02'}); however, instead
  of (\ref{r(x)}), we choose the following function $r(x)$:
\bear                                                 \label{r>}
     r^2(x) = (|x| + 1)^2 -1, &&  |x| > d,
\\                                                    \label{r<}
     r^2(x) = ax^2 + s = \frac{d+1}{d}x^2 +d, && |x| < d,
\ear
  where $d  = \const > 0$ is sufficiently small and the constants in
  (\ref{r<}) are chosen to make $r$ and $r'$ continuous at $x = \pm d$,
  as shown in Fig.\,3.

\begin{figure}
    \begin{center}
\fbox{\includegraphics[ width=8.2cm,height=8cm,]{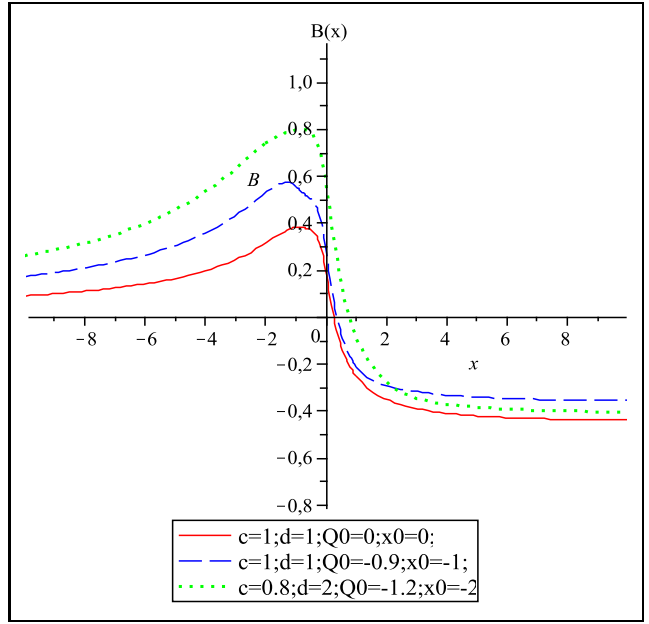}}
    \end{center}
\caption{The function $B(x)$ obtained from a smoothed but peaked $f(u)$}
\hrulefill\
\end{figure}         % fig4

\begin{figure}
    \begin{center}
\fbox{\includegraphics[ width=8.2cm,height=8cm,]{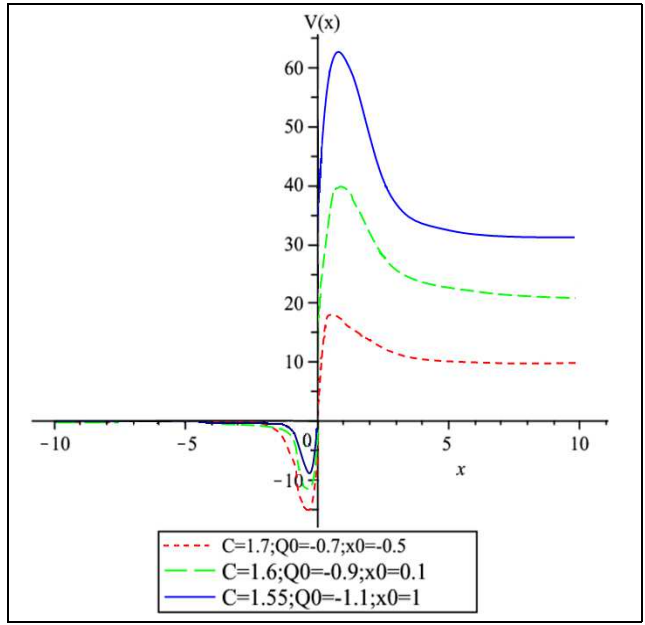}}
    \end{center}
\caption{The potential $V(x)$ obtained from a smoothed but peaked $f(u)$}
\hrulefill\
\end{figure}         % fig5

  Let us take $f(x) \equiv 0$ at $|x| > d$ and, for $|x| < d$, as in
  (\ref{f1}), $f = 2C/r^6$. Then from (\ref{cons}) we find $Q(x)$ as follows:
\beq    \nq
      Q(x) = \vars{  Q_0, &  x < -d;\\
             Q_0 + [B(x) - B(-d)] C, & |x| \leq d;\\
             Q_0 + [B(d) - B(-d)] C, & x > d,
          }
\eeq
  where $Q_0$ is an integration constant. The constant $C$ (more precisely,
  its dimensionless counterpart ${\bar C} = C/b^4$) is determined from \eq
  (\ref{01'}), where we require continuity of $\psi'$ at $x= \pm d$. \eq
  (02') for the function $B(x) = A/r^2$ is then solved analytically for
  $|x| > d$ but only numerically for $|x| < d$.

  As a result, at $x \leq -d$ we obtain $B(x)$ in the form (\ref{B-}) with
  $c = 1$. At $x \geq d$, an expression for $B(x)$ is similar to (\ref{B+})
  but obtained with initial conditions for $B$ and $B'$ at $x=d$ that follow
  from numerical integration of \eq (\ref{02'}) in the range $|x| < d$.
  The results are shown in Fig.\,4. Fig.\,5 shows the corresponding
  potential $V(x)$. Clearly these are black-universe models, where the
  asymptotic behavior of $V(x)$ approaching a positive constant as
  $x\to\infty$ corresponds de Sitter expansion with a positive cosmological
  constant.

\section{Concluding remarks}

  We have built a family of black-universe solutions to the modified Einstein
  equations valid in an RS2 type brane world. They have been obtained without
  explicitly introducing any phantom matter. Just as was the case with \wh\
  solutions \cite{bwh1,bwh2}, the role of exotic matter in the field
  equations is played by the ``tidal'' term of geometric origin, which
  has no reason to respect the energy conditions known for physically
  plausible matter fields.

  This new kind of solutions, having a \bh\ nature as seen from large
  negative $x$, supplements the sets of known examples of both \sph\ \bwd\
  \bhs\ (see, e.g., \cite{bw-bh, bbh1}) and \bhs\ with scalar``hair'' (see,
  e.g., \cite{pha1, pha4, sca-bh}).

  Let us recall that the existence of black-universe models suggests the idea
  that our Universe could appear from another, ``mother'' universe and
  undergo isotropization (e.g., due to particle creation) soon after crossing
  the horizon. The \KS\ nature of our Universe, as opposed to the more
  popular spatially flat models, is not excluded observationally \cite{craw}
  if its isotropization had happened early enough, before the last scattering
  epoch (at redshifts $z\gtrsim 1000$). A tentative estimate obtained
  for cosmological evolution beginning from a horizon (the so-called Null Big
  Bang) in another kind of model, that with a vacuumlike static core, has
  shown \cite{bd07} that such models isotropize very quickly and can now be
  quite observationally indistinguishable from isotropic ones. One can also
  notice that we are thus facing one more mechanism of universes
  multiplication, in addition to the well-known mechanism existing in the
  chaotic inflation scenario.

  The presently obtained models do not pretend to be quite realistic, they
  simply show the possibility of such a scenario in principle. As any
  solutions to the effective 4D equations describing gravity on the brane,
  they certainly need extension to the bulk, whose finding is quite a
  challenging problem, although an extension always exists due to the known
  embedding theorems. Another problem to be solved is that of stability of
  such solutions.

\Acknow
   {This work was supported in part by the Russian Foundation for
    Basic Research grant 09-02-0677-a and by a grant of People Friendship
    University of Russia (NPK MU).}

\end{document}